\documentclass{article}

\usepackage{theorem}

\theorembodyfont{\normalfont}

\usepackage{diaginca}

\begin{document}

\author{Jean-Guillaume Dumas\footnote{
   Universit\'e de Grenoble, laboratoire de mod\'elisation
   et calcul, LMC-IMAG BP 53 X, 51 avenue des math\'ematiques,
   38041 Grenoble, France.
   \texttt{\{Jean-Guillaume.Dumas,~Dominique.Duval\}@imag.fr}
.}~~and Dominique Duval\footnotemark[2]}
\title{Towards a diagrammatic modeling of 
the \linbox C++ linear algebra library\footnote{supported by
  the Institut d'Informatique et de Math\'ematiques Appliqu\'ees de
  Grenoble, InCa project.}}
\date{\today}
\maketitle
\begin{abstract}
We propose a new diagrammatic modeling language, DML.
The paradigm used is that of the category theory and in particular of
the pushout tool. We show that most of the object-oriented
structures can be described with this tool and have many examples in
C++, ranging from virtual inheritance and polymorphism to template
genericity.
With this powerful tool, we propose a quite simple description of the C++
\linbox library. This library has been designed for efficiency and
genericity and therefore makes heavy usage of complex template and
polymorphic mechanism. Be reverse engineering, we are able to
describe in a simple manner 
the complex structure of archetypes in \linbox.
\end{abstract}

\section{Introduction}
The \linbox library is a C++ template library for exact,
high-performance linear algebra computations with dense, sparse, and
structured matrices over the integers and over finite fields.
C++ templates are used to provide both high performance and
genericity \cite{jgd:2002:icms}. 
In particular, \linbox algorithms are generic with respect to the field
or ring over which they operate and with respect to the internal 
organization of the black box matrix. \linbox aims to provide
world-class high performance implementations of the most advanced
algorithms for exact linear algebra. Combining this high-performance
and the genericity resulted in an intricate system of C++ classes.

In this paper, we propose a reverse engineering of this system in
order to enlight its underlying mechanism and describe its various
functionalities in a unified way. The chosen paradigm is that of
\emph{diagrammatic modeling} and \emph{categories}.
Our major categorical tool is the notion of \emph{pushout},
which corresponds to several constructions in C++. 
Pushouts are widely used for describing the combination of two 
specifications sharing a common part, see for example 
\cite{Goguen:1973:CFG,Srinivas:1995:SPEC,Burstall:1997:PTT,
Oriat:2000:DEM}.
However, diagrammatic modeling languages
like UML \cite{Muller:2000:UML} are inadequate for dealing 
with such pushout constructions.
For instance, in UML class diagrams and object diagrams
are distinct, while we propose to merge them into a unique 
kind of diagram, where an instantiation (between a class and an object)
is at the same level as an association between two classes
or a link between two objects. 
Moreover, unlike UML, we consider the relation between a 
generic class and its template parameter as a kind of association, 
which allows to consider parameter passing as a pushout construction.
Therefore, we propose to use a new diagrammatic 
modeling language (called DML), significantly different from UML. 
In particular, we identify the object-oriented notions of parameter
passing, virtual inheritance, polymorphism template parameter passing, 
object instantiation, as pushout constructions in a category.
It should be noted that many arrows, for example the inheritance arrows,
are directed in the opposite way in UML and in DML.  

Some basic features about categories are used in
this paper; they can be found in many textbooks, like
\cite{MacLane:1997:CWM,Barr:1990:CTCS}. 
They are also given here, in section~\ref{sec:pushout},
for the sake of completeness. Then we present in section \ref{sec:dml}
the new diagrammatic modeling language DML.
Since \linbox is a C++ library, this presentation of DML is based on
the object-oriented language C++ (and Java for a while), 
but it could be adapted to another object-oriented language. 
Finally DML is used for analyzing part of the structure of the 
LinBox C++ library in section~\ref{sec:linbox}.



\section{Categories and pushouts}
\label{sec:pushout}

\subsection{Categories}
\label{ssec:cat}

A category can be seen as a generalized monoid.
For example, 
let $\cF$ denote the functions on the reals, i.e., the functions 
from $\bR$ to $\bR$, like $\sin, \cos, \exp:\bR\to\bR$.
Such functions can be composed, like $\exp.\sin$ 
(or $\exp\circ\sin$), defined by $\exp.\sin(x)=\exp(\sin(x))$.
This yields a structure of \emph{monoid} on the set $\cF$:
this means that the composition is associative,
i.e., $(f.g).h=f.(g.h)$, which is therefore denoted $f.g.h$,
and that there is a unit for the concatenation,
namely the identity $\id$, defined by $\id(x)=x$, 
such that $f.\id=f$ and $\id.f=f$.

Now, let $\cF$ denote the functions from $X$ to $Y$,
where $X$ and $Y$ can be either $\bR$ or $\bC$,
for instance $\bR\to\bR$ (sine function), 
$\bC\to\bC$ (complex conjugate), 
$\bC\to\bR$ (modulus) or $\bR\to\bC$ (inclusion). 
Such functions can still be composed, 
but only if they are consecutive:
if $f:X\to Y$ and $g:Y\to Z$, then $g.f:X\to Z$.
The associativity axiom is still valid, when it makes sense. 
There are now two identities, $\id_{R}: \bR\to\bR$
and $\id_{C}:\bC\to\bC$. The unitarity axioms 
become: if $f:X\to Y$ then $f.\id_X=f$ and $\id_Y.f=f$.
This is no more a structure of monoid,
because of the typing restrictions, but a structure of \emph{category},
as defined below.

\begin{defn}
A \emph{category} $\cC$ is made of 
\emph{points} $X$, $Y$,\dots
and \emph{arrows} $f$, $g$,\dots,
each arrow has a  \emph{source} and a \emph{target}
(this is denoted $f:X\to Y$ or $X\uto{f}Y$), 
each point $X$ has an \emph{identity} arrow $\id_X:X\to X$,
each pair of consecutive arrows $X\uto{f}Y\uto{g}Z$
has a \emph{composed} arrow $X\uto{g.f}Z$, 
and moreover the associativity and unitarity axioms are satisfied: 
$(h. g). f = h. (g. f)$, $f.\id_X = f $ and $\id_Y. f = f$,
as soon as it makes sense. 
\end{defn}

\subsection{Inheritance}
\label{ssec:herit}
A category can also be seen as a generalized ordering.
For example, let us look at the inheritance relation 
in an object-oriented language. 
When multiple inheritance is forbidden, as in Java, 
the inheritance relation defines a partial order on classes:
if $Z$ inherits from $Y$, which inherits  from $X$,
then, by transitivity, $Z$ inherits from $X$. 
Let us introduce an arrow $X\to Y$ whenever $Y$ inherits from $X$.
\textbf{Warning!} 
\textsl{This is the opposite of the notation that can be found in most
diagrammatic approaches, e.g. in UML or in \cite{Stroustrup:1997:cpp};
the reasons for this choice will be exposed 
in section~\ref{sec:dml}.}
Now, the transitivity of the inheritance relation 
corresponds to the composition of arrows:
if there are two consecutive arrows $X\to Y\to Z$,
then there is a composed arrow $X\to Z$.
It is not necessary to give a name to the arrows,
since there is at most one arrow with given source and target.

In an object-oriented language 
that does allow multiple inheritance, 
two situations may occur, they are called 
respectively the \emph{ordinary inheritance} and the 
\emph{virtual inheritance} in C++ \cite{Stroustrup:1997:cpp}.
If $X\to Y_1\to Z$ and $X\to Y_2\to Z$,
in the ordinary inheritance relation 
$Z$ inherits from $X$ in two different ways,
while in the virtual inheritance relation
$Z$ inherits from $X$ in only one way.
Let us give a name to the inheritance arrows:
$X\uto{f_1} Y_1\uto{g_1} Z$ and $X\uto{f_2} Y_2\uto{g_2} Z$.
From a categorical point of view, 
there are two composed arrows 
$g_1.f_1:X\to Z$ and $g_2.f_2:X\to Z$.
If nothing more is said, they are distinct, 
which corresponds to the ordinary inheritance.
But if the equality $g_1.f_1=g_2.f_2$ is added, 
this corresponds to the virtual inheritance.

From now on, using C++ terminology, 
we say that a {\em derived} class (or subclass) 
inherits from a {\em base} class (or superclass).
The base class may be {\em abstract}: it is a class with
pure virtual methods in C++, or an \verb!interface! in java;
the idea is to provide an interface
that the derived classes have to follow (mandatory methods);
this kind of inheritance is used in section~\ref{ssec:polym}. 
Inheritance can also be an {\em extension}, 
where the derived class adds new functionalities or members 
to the base class, see e.g. \cite{Taivalsaari:1996:NI} 
for more details on inheritance. In both
cases, anyway, the derived class adds something: 
in the first case, only implementations are added.

\subsection{Pushouts}
\label{ssec:push}

Let $\cC$ be a category.
A \emph{span} $\Sp$ in $\cC$ is made of two arrows with a common source:
  $$
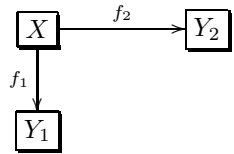
\xymatrix@C=4pc{
    *+[F-,]{X} \ar[d]_{f_1} \ar[r]^{f_2}  & *+[F-,]{Y_2} \\
    *+[F-,]{Y_1} &  \\
  } $$
A \emph{cone} $\Co$ with \emph{base} $\Sp$ in $\cC$ 
is made of the span $\Sp$ together with 
a point $Z$, called the \emph{vertex} of $\Co$,
and two arrows $g_1:Y_1\to Z$, $g_2:Y_2\to Z$, 
called the \emph{coprojections} of $\Co$, 
such that $g_1.f_1=g_2.f_2$: 
  $$
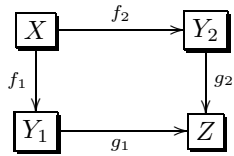
\xymatrix@C=4pc{
    *+[F-,]{X} \ar[d]_{f_1} \ar[r]^{f_2}  & *+[F-,]{Y_2} \ar[d]^{g_2} \\
    *+[F-,]{Y_1} \ar[r]_{g_1} & *+[F-,]{Z} \\
  } $$

The pushout of a span $\Sp$ is defined below
as a cone with base $\Sp$ 
which satisfies some \emph{initiality} condition
(i.e., some kind of ``minimality'' condition).
In this paper, the coprojections of a pushout cone are 
represented as dashed arrows.
\begin{defn}
A \emph{pushout} with base $\Sp$ is a cone $\Co$
with base $\Sp$ such that, for each cone $\Co'$ 
with the same base $\Sp$, there is a unique arrow $h:Z\to Z'$ 
such that $h.g_1=g'_1$ and $h.g_2=g'_2$:
   $$
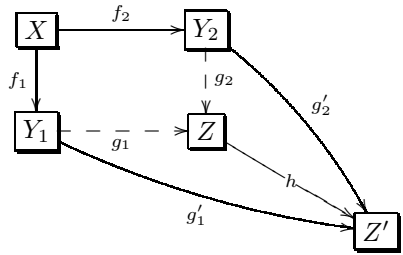
\xymatrix@C=4pc{
    *+[F-,]{X} \ar[d]_{f_1} \ar[r]^{f_2}  & *+[F-,]{Y_2} \ar@{-->}[d]^{g_2} 
       \ar@/^/[ddr]^{g'_2} & \\
    *+[F-,]{Y_1} \ar@{-->}[r]_{g_1} \ar@/_/[drr]_{g'_1} & 
      *+[F-,]{Z} \ar[dr]|{h} & \\
     && *+[F-,]{Z'}  \\
  } $$
\end{defn}
A span $\Sp$ cannot have more than one pushout (up to isomorphism),
which is called \emph{the} pushout with base $\Sp$.
The point $X$ will be called the \emph{gluing point} of $\Sp$.

Roughly speaking, 
this means that $Z$ is obtained by ``gluing 
$Y_1$ and $Y_2$ along the image of $X$''.



\section{DML: a Diagrammatic Modeling Language}
\label{sec:dml}

\subsection{The Category for DML}
\label{ssec:cdml}

In order to model the structure of a C++ piece of 
software, a category $\cdml$ is described now,
in a rather informal way;
a more precise definition of the category $\cdml$
would deserve a longer paper.

The points of the category $\cdml$ are called the \emph{specifications}. 
They are, essentially, the C++ \emph{types}.
More precisely, a specification may correspond to 
a built-in type, a class or a typename, 
and it may also correspond to 
a value in a built-in type or an instance of a class.
Essentially, a specification $A$ is seen as a collection of \emph{members},
and it determines a set of \emph{models} $\Mod(A)$.
If the specification is a class $A$, 
its models are the instances of the class $A$.
If it is an object $a$, its unique model is itself.
So, one may look at a specification 
either from a \emph{syntactic} point of view, i.e., 
as a collection of members, 
or from a \emph{semantic} point of view, i.e., 
as a set of models.
In this paper, we use the syntactic point of view.

The arrows of the category $\cdml$ 
are the morphisms between the specifications,
they are of various kinds. 
Since we use the syntactic point of view on specifications, 
a morphism $\varphi:A\to B$ 
maps each member of $A$ to a member of $B$
(or to some composition of members of $B$).
For example, a morphism of specifications can be an 
\emph{inheritance} morphism, between two classes. 
When $B$ inherits from $A$,
the class $B$ contains all the members of the class $A$,
plus some new ones. From the syntactic point of view, 
inheritance is an arrow $\varphi:A\to B$.
This morphism induces a map $\Mod(\varphi):\Mod(B)\to\Mod(A)$, 
by omitting the interpretation of the members of $B$
that are not members of $A$.
For this reason, it is often illustrated as an arrow \emph{from $B$ to $A$}:
this is the case in UML, for instance,
but in this paper the syntactic orientation $\varphi:A\to B$ 
is always chosen.  
A \emph{template parameterization} is also
a morphism of specifications.
When a template class $T$ occurs as 
a template parameter for a class $B$,
the members of $T$ can be used in $B$,
so that there is a morphism $T\to B$.
An \emph{instantiation} is another kind of 
morphism of specifications.
When an object $a$ is created as an instance of a class $A$,
then the members of $A$ are instantiated in $a$, 
which can be seen as a morphism $A\to a$.
An \emph{implementation} of an abstract class $A$
by a class $B$ is also a morphism $A\to B$.

Specifications may be built progressively, by systematic constructions, 
thanks to pushouts.
The aim of the next subsections is to show
that some pushouts in the category $\cdml$ 
correspond to fundamental constructions in C++: 
virtual inheritance is detailed in section \ref{ssec:vInherit},
and standard parameter passing in section \ref{ssec:param};
now object oriented polymorphism is described in section \ref{ssec:polym}, 
template parameter passing in section~\ref{ssec:template},
and object instantiation in section~\ref{ssec:instance}.
Several examples, from the library \linbox,
are given in section~\ref{sec:linbox}.

\subsection{Virtual inheritance}
\label{ssec:vInherit}
Virtual inheritance gives rise to cones, as explained in section~\ref{ssec:herit}.
Moreover, such a cone is a pushout if and only if the doubly derived class $Z$
is ``minimal'', in the sense that $Z$ has no additional member, 
on top of those that are inherited from $Y_1$ and $Y_2$.
The corresponding piece of C++ code, when the methods are neither
constructors nor destructors, is as follows:
\begin{verbatim}
struct X {void m0(){...} }; 
struct Y1: public virtual X {void m1(){...} }; 
struct Y2: public virtual X {void m2(){...} }; 
struct Z: public virtual Y1, public virtual Y2 { }; 
\end{verbatim}
Then the methods $m_1$, $m_2$,  and one 
method $m_0$, are inherited by $Z$. 
When some $m_i$ is a constructor, since it is not inherited in C++, 
it must appear explicitly in the derived class. 
We still speak of
pushout in the latter, despite this adjunction.

\subsection{Parameter passing}
\label{ssec:param}

The formalization of multiple inheritance
by a pushout, as above, 
is an example of a \emph{symmetric} use of pushouts,
where both arrows in the span are of the same nature.
In this paper, we are rather interested by several kinds of 
\emph{dissymmetric} ways to use pushouts \cite{Barr:1990:CTCS}. 
The paradigm for the constructions in the next sections 
is the parameter passing construction, as described now.

\newcommand{\Fxa}{\begin{xy}
  \xymatrix@R=1pc{
    X \ar[r]^{f} & Y \\ 
     }  
  \drop++\frm<20pt>{-}
  \end{xy}}
\newcommand{\fXa}{\begin{xy}
  \xymatrix@R=1pc{
    X \\ 
     }  
  \drop++\frm<20pt>{-}
  \end{xy}}
\newcommand{\fxA}{\begin{xy}
  \xymatrix@R=1pc{
    U \ar[r]^{a} & X \\ 
     }  
  \drop++\frm<20pt>{-}
  \end{xy}}
\newcommand{\FxA}{\begin{xy}
  \xymatrix@R=1pc{
    U \ar[r]^{a} & X \ar[r]^{f} & Y \\ 
     }  
  \drop++\frm<20pt>{-}
  \end{xy}}
\newcommand{\FxAd}{\begin{xy}
  \xymatrix@R=1pc{
    U \ar[r]^{a} \ar@/_2ex/[rr]_{f.a}  & X \ar[r]^{f} & Y \\ 
    \mbox{} \\ 
     }  
  \drop++\frm<20pt>{-}
  \end{xy}}
\newcommand{\FA}{\begin{xy}
  \xymatrix@R=1pc{
    U \ar[r]_{f.a}  & Y \\ 
     }  
  \drop++\frm<20pt>{-}
  \end{xy}}
\newcommand{\dashdownarrow}{\begin{xy}
  \xymatrix{
    \ar@{-->}[d] \\
  \\
     } 
  \end{xy}}
\newcommand{\longdownarrow}{\begin{xy}
  \xymatrix{
    \ar[d] \\
  \\
     } 
  \end{xy}}
Given some expression $f(x)$ and some value $a$ for $x$,
the parameter passing construction 
builds the expression $f(a)$. 
Here $f:X\to Y$ is a function,
$x:X$ is a symbol called the \emph{formal parameter}, 
and $a:X$ is a constant called the \emph{actual parameter},
so that the result $f(a)=f.a:Y$ is also a constant.
The parameter and the result are considered as 
constant functions $a:U\to X$ and $f(a)=f.a:U\to Y$,
where $U$ is the \emph{unit} type, 
which is interpreted as a singleton. 
So, the parameter passing process 
is seen as an application of the rule for 
the composition of arrows: 
$$ \frac {U\uto{a}X \qquad X\uto{f}Y} {U\uto{f.a}Y} $$
The pushout construction is used for building 
an instance of the premises of this rule.
The category where the pushout takes place is
the category $\cG$ of (directed multi-)graphs.
First the data, made of $f:X\to Y$ and $a:U\to X$,
with the same type $X$, is represented as a span $\Sp$
in the category $\cG$:
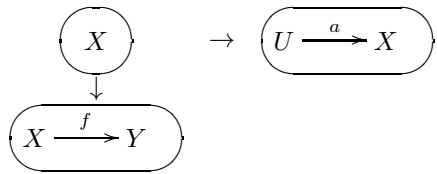
$$ \begin{matrix}
\fXa & \rightarrow & \fxA \cr 
\downarrow & & \cr
\Fxa & & \cr
\end{matrix}$$
Then the pushout with base $\Sp$ is built
in the category $\cG$:
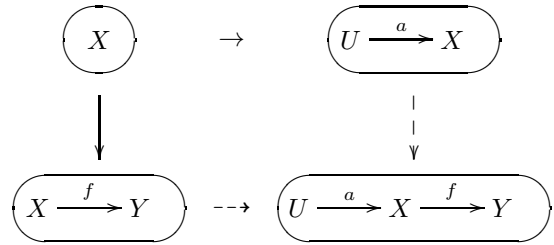
$$ \begin{matrix}
\fXa & \rightarrow & \fxA \cr 
\longdownarrow & &\dashdownarrow  \cr
\Fxa & \dashrightarrow & \FxA \cr
\end{matrix}$$

The vertex of the pushout is an instance of 
the premises of the rule for composition.
Then, the constant $f.a$ is obtained by applying this rule.
More about this point of view on deduction rules 
can be found in \cite{Duval:2002:DS,Duval:2003:DS}. 

So, schematically, the parameter passing 
construction corresponds to the following pushout:
\voir{here, $x$ reappears.}
  $$
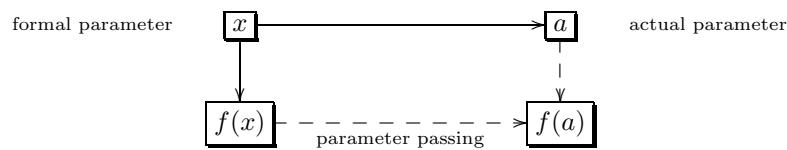
\xymatrix@C=8pc{
    \mbox{} \ar@{}[r]|{\text{formal parameter}} &
    *+[F-,]{x} \ar[d] \ar[r] & 
      *+[F-,]{a} \ar@{-->}[d] &
      \mbox{} \ar@{}[l]|{\text{actual parameter}}  \\
    & *+[F-,]{f(x)} \ar@{-->}[r]_{\text{parameter passing}} & 
      *+[F-,]{f(a)} & \\
  } $$

\subsection{Template parameter passing}
\label{ssec:template}

In C++, a class can be used as a parameter
for building a new class,
thanks to the \emph{template} parameters mechanism.
This works just like classical parameters, i.e., 
template parameter passing can be formalized 
as a pushout of specifications:
  $$
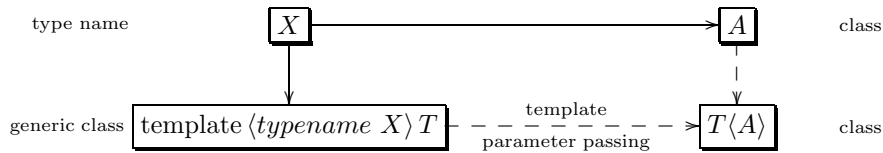
\xymatrix@C=8pc{
    \mbox{} \ar@{}[r]|{\text{type name}} &
    *+[F-,]{X} \ar[d] \ar[r] & 
      *+[F-,]{A} \ar@{-->}[d] &
      \mbox{} \ar@{}[l]|{\text{class}\qquad}  \\
    \mbox{} \ar@{}[r]|{\text{generic class}\quad} &
    *+[F-,]{\text{template}\,\tpl{typename~X}\,T} 
      \ar@{-->}[r]^{\quad\quad\quad\quad\text{template}}_{\quad\quad\quad\quad\text{parameter passing}} & 
      *+[F-,]{T\tpl{A}} & 
      \mbox{} \ar@{}[l]|{\text{class}\qquad}  \\
  } $$
Here, $X$ is the name of the formal template parameter,
that is used in the definition of the generic class $T$.
The class $A$ is the actual value to be passed, 
and the vertex of the pushout is the resulting class 
$T\tpl{A}$.

\subsection{Object instantiation} 
\label{ssec:instance}

Object instantiation can be obtained from a pushout of specifications, 
in many different ways.
For instance, the morphism above, 
from the class $A$ to the class $T\tpl{A}$, 
can be used for building an instance of $T\tpl{A}$
from an instance of $A$, as follows:
  $$
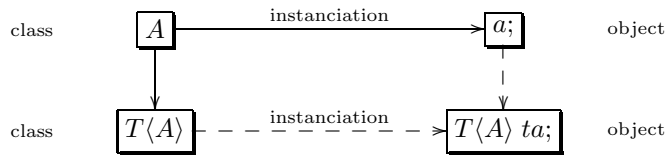
\xymatrix@C=8pc{
    \mbox{} \ar@{}[r]|{\qquad\text{class}} &
    *+[F-,]{A} \ar[d] \ar[r]^{\text{instanciation}} & 
      *+[F-,]{a;} \ar@{-->}[d] &
      \mbox{} \ar@{}[l]|{\text{object}\qquad}  \\
    \mbox{} \ar@{}[r]|{\qquad\text{class}} &
    *+[F-,]{T\tpl{A}} 
      \ar@{-->}[r]^{\text{instanciation}} & 
      *+[F-,]{T\tpl{A}\;ta;} & 
      \mbox{} \ar@{}[l]|{\text{object}\qquad}  \\
  } $$
This pushout yields an instance $ta$ of $T\tpl{A}$ 
using the empty constructor of T, which might call constructors of $A$.

\subsection{Object oriented polymorphism}
\label{ssec:polym}
In this section we propose to see the polymorphism mechanism of an
object oriented language, involving inheritance, as a pushout.
According to \cite[\S 12.2.6]{Stroustrup:1997:cpp}, object oriented polymorphism
behaves as follows in C++: 
the member functions called must be \emph{virtual} 
and objects must be manipulated through pointers 
or references. 
So, polymorphism is obtained when a derived class is used 
via a pointer to its base class, 
and when this base class contains virtual member functions,
for example it can be an abstract class. 
The idea is to write
algorithms on the base class and pass afterwards derived
values. 
Note that the effect is  close to that of the \verb!template! 
mechanism of C++.
Here is a C++ example : 

\begin{verbatim}
#include <iostream>
struct A { virtual void f() = 0; }; // abstract class
void g(A * a) { a->f(); }           // global function

// derived class adds an implementation of method f.
struct B : public A { void f() { std::cout << "f of B"; } };

int main() {
    B b;       
    g( &b );  // a pushout is used 
    return 0;
}
\end{verbatim}

In this example, the class $A$ is an abstract class,
with a virtual method $f$;
on the one hand, the class $B$ inherits from $A$,
and adds an implementation of $f$;
on the other hand, 
the function $g$ is implemented knowing only the interface of $f$,
as given in $A$.  
Later on, a pointer on an object $b$ of type $B$
can be passed as an argument to this function $g$.
The corresponding pushout 
is then between $A$ and $g$ on one side, 
and the derived class $B$ on the other side, 
with the abstract class $A$ as gluing point:
  $$
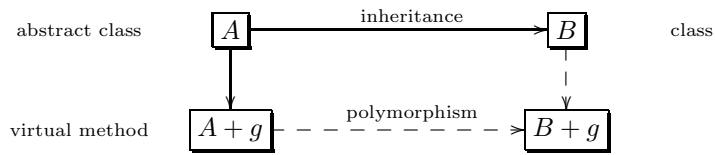
\xymatrix@C=8pc{
    \mbox{} \ar@{}[r]|{\text{abstract class}} &
    *+[F-,]{A} \ar[d] \ar[r]^{\quad\text{inheritance}} & 
      *+[F-,]{B} \ar@{-->}[d] &
      \mbox{} \ar@{}[l]|{\text{class}\qquad}  \\
   \mbox{} \ar@{}[r]|{\text{virtual method}} &
    *+[F-,]{A+g} 
      \ar@{-->}[r]^{\quad\text{polymorphism}} & 
      *+[F-,]{B+g} & 
  \\
  } $$

\subsection{A C++ example} 
\label{ssec:cplusplus}
The next C++ piece of code provides an example 
of template parameter passing followed by 
object instantiation.

\begin{verbatim}
#include <iostream>
template <typename X> struct T { // Template class
    // All X class are supposed to have a "g" method
    // The code in T, defines the required interface
    void f(X x) { x.g(); }       
};

// An actual implantation matching the "X" interface
struct A {
    void g() { std::cout << "g of A"; }
};

int main() {
    T<A> ta;   // The class A is given as a parameter of T
    ta.f();
    return 0;
}
\end{verbatim}


\section{Application to a linear algebra C++ library: \linbox}
\label{sec:linbox}

\linbox is a C++ template library of routines 
for solving linear algebra problems.
It has been designed for dealing 
with matrices over a variety of domains,
in particular over finite fields or rings.
Genericity and high performance are the twin goals of
\linbox.  
The genericity is achieved by use of a small set of interfaces.  
Algorithms are implemented with C++ template parameters 
which may be instantiated with any class adhering to the
specified interface.
High performance is achieved by judicious specializations of the
generic algorithms.
It is entirely within the spirit of the project to introduce new
implementations. 
Thus a user of the library may invoke a \linbox algorithm, 
say for determinant or rank of a matrix, 
but providing a black box class of her own design and
perhaps even providing the underlying field 
(or commutative ring) representation.
Conversely, the \linbox~field and ring interfaces and the many
specific representations can be used for purposes other than linear
algebra computation or with algorithms not provided by \linbox.

In order to solve simultaneously the genericity and performance issues, 
the \linbox library has designed a complex structure,
involving five distinct classes, that each \linbox domain must respect.
This system is extremely efficient \cite[Table
5.1]{Turner:2002:these}, but it is quite complex to describe and to use. 
The aim of this paper is to give an abstract view on 
this architecture, 
thanks to the Diagrammatic Modeling Language,
in order to get a simple user interface description. 
We now focus on the case where the underlying domain is a field,
but a similar structure holds for commutative rings, for example.

\subsection{A class ``Field'' for the algorithms}
\label{ssec:code}
\linbox algorithms have been conceived to function with a variety of fields,
in particular over finite fields or rings.
To carry out its computations, any algorithm may need additional
parameters, such as the modulus \code{p} (a prime number)
for computing in the field of integers modulo \code{p}.
For this purpose, \linbox offers a special object for the field 
which defines its methods (e.g., the arithmetic operators). 
A \code{Field} class thus contains the actual computational code. 
An instance \code{F} of the class \code{Field} corresponds to 
an actual field with its parameters.
Moreover, an internal class \code{Field::Element} 
is used to deal with the storage of the elements 
of this field:
for example, the call \code{F.add(x, y, z)}
adds the elements \code{y} and \code{z} in the field \code{F} and
stores the result in the element \code{x}. 
This \code{Field::Element} type can be a \code{long int} of C++, 
for the field of integers modulo \code{p}
when \code{p} is a word size prime, 
or it can be a more complex data structure.

\subsection{An abstract class ``FieldAbstract'' for genericity}
\label{ssec:abstract}
Since all the algorithms are generic
with respect to the field, they must follow a common interface. 
This is carried out in \linbox by a common inheritance to an abstract class, 
\code{FieldAbstract}.

\subsection{An archetype ``FieldArchetype'' class to control code bloat}
\label{ssec:archetype}
The number of generic levels in \linbox induces the need
to control code bloat. The solution developed by \linbox
is to define a non generic additional class, \code{FieldArchetype}
\cite{jgd:2002:icms}.
This class encloses a pointer towards a \code{FieldAbstract} object. 
The generic algorithms can thus be instantiated on this single class.
If the explosion of executable code makes it necessary, they are
thus linearly and separately compiled. Thus, it is not
directly the abstract class which plays the interface part. 
This prototype \cite{Turner:2002:these, kaltofen:2005:memory} fulfills 
three roles in the library:
it describes the common object interface, 
it provides a compiled instance of executable code, 
and it controls code bloat.
Separating this archetype from the abstract class is mandatory in the
field case for efficiency. Indeed, while polymorphism could be
directly used, it induces the overhead dereferencing the pointers.
This dereferencing can be too much a price to pay for every
arithmetic operation. Thus, in \linbox, one can choose between best
efficiency without archetypes, or better control of code bloat to the
price of an overhead in computational timings.

\subsection{A generic envelope class ``FieldEnvelope'' for external fields}
\label{ssec:envelope}
Two problems result from this organization. 
First of all, every field, even if comes from an external library, 
must inherit from the abstract class. 
Second, the constructors, and in particular the copy constructor, 
cannot be inherited in C++. It is thus necessary to add a virtual
method \code{clone} fulfilling this job in the abstract class. 
This induces a different interface between the abstract class and the
archetype class. 
These two problems of inheritance and interface are
solved by the creation of an additional wrapping class
\code{FieldEnvelope}, which we describe precisely now.
Then, for every field, its envelope inherits from the 
abstract class \code{FieldAbstract}.

An envelope is a generic adaptor \cite{Wise:1996:stl}, 
matching the interface of its wrapped object.
In C++, we distinguish between several variants of this design.
All of them are templated structures, 
depending on a template parameter $B$.
\begin{enumerate}
\item \emph{Envelope without inheritance:}
the envelope class is related to $B$ via the type 
of a member, either directly or via a pointer. 
Since there is  no inheritance, 
object oriented polymorphism is impossible.
\begin{enumerate}
\item \emph{Copy envelope:}
  the template parameter $B$ is the type of a member
  of the envelope class. 
\begin{verbatim}
template <typename B> struct Env {
private:
   B _b;
};
\end{verbatim}
\item \emph{Pointer envelope:} 
  this is a variant of the latter, without copy:
  it is $B*$, instead of $B$, which is the type of a member
  of the envelope class. 
\begin{verbatim}
template <typename B> struct Env {
private:
   B* _b;
};
\end{verbatim}
\end{enumerate}
\item \emph{Inheritance envelope:} the object inherits from its
  template. Every template characteristic is preserved except for the
  constructors and destructors. 
  Object-oriented polymorphism is also preserved.
\begin{verbatim}
template <typename B> struct Env : public B;
\end{verbatim}
\end{enumerate}

The envelope fulfills two functionalities.
On the one hand, an envelope gives an internal inheritance 
to immutable external classes: 
with multiple inheritance, an immutable class can 
nonetheless use polymorphism.
\begin{verbatim}
// Every Env<B> class inherits from a class A.
template <typename B> struct Env : public B, public A;
\end{verbatim}
On the other hand, an envelope allows generic method abstraction.
Indeed, it is conceptually impossible
to define an abstract class with templated methods:
the virtual table mechanism cannot resolve an 
associated abstract method call,
since the actual method is not known when the object is created.
The envelope mechanism can simulate an abstract class 
by forcing the template parameter interface. 
\begin{verbatim}
template <typename B> struct Env : public B {
   template <typename V> void mygenericmethod(V a) {
        // This method is required for every B even if it is generic
        B::mygenericmethod(a);
   }
};
\end{verbatim}

\subsection{Envelopes in Java}
\label{ssec:java}

This envelope formalism is not restricted to C++.
In Java, for instance, one can build similar classes.
First, an external and independent class:
\begin{verbatim}
public class External {
        int _a;
        External(int a) {
                _a = a;
        }
        public void amethod() {
        System.out.println("a: " + _a);
        }
}
\end{verbatim}
Then an abstract class:
\begin{verbatim}
public interface Abstract{
        public void Themethod();
}
\end{verbatim}
And finally an envelope \code{EnvelopeInherit} 
that implements \code{Abstract} and extends \code{External}:
\begin{verbatim}
public class EnvelopeInherit extends External implements Abstract {
        EnvelopeInherit(int a) {
                super(a);
        }
        public void Themethod() {
                super.amethode();
        }
}
\end{verbatim}
The \code{EnvelopeInherit} class is therefore 
an \code{External} class implementing
the internal \code{Abstract} interface. 
The copy envelope is defined similarly. 
Now it is less impressive in Java, since the envelope cannot
be templated. 
In C++ a single envelope can be used by many external classes.

\subsection{A DML description of the \linbox architecture for domains} 
\label{ssec:specif}

\begin{figure}[htbp]
\input{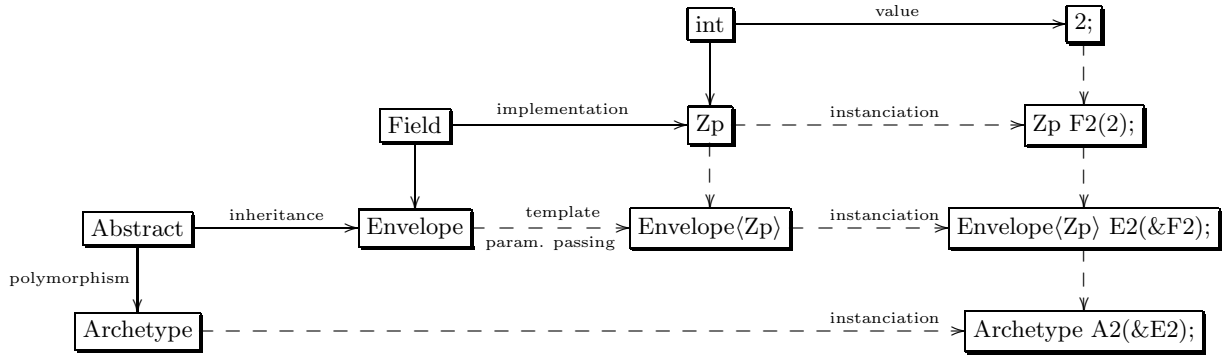}
\caption{A DML diagram for the \linbox architecture for fields,
  with a copy envelope}\label{fig:corpsrecop}
\end{figure}

In order to visualize the relations between the various classes 
that are used by \linbox for representing fields, 
we use the Diagrammatic Modeling Language of section~\ref{sec:dml}.
The specifications that are used by \linbox
for dealing with one field (here the field with two elements),
except for the class \code{Field::Element}, 
are represented by the diagram in figure~\ref{fig:corpsrecop}.

This diagram can be read from the left to 
the right, i.e., from the abstraction to the instances.
With the exception of the first line, 
the specifications (i.e., the boxes) 
in the rightmost column are objects,
the other specifications are classes 
(\code{Zp} denotes a class of finite prime fields),
and the rightmost horizontal arrows are instantiations.
The first line is slightly different: 
it has a built-in type \code{int} instead of a class, 
and a value \code{2} of type \code{int} instead of an instance.
The three objects \code{F2}, \code{E2} and \code{A2}
represent the field with two elements,
from three different points of view.
The three pushouts on the right, 
with vertices \code{F2}, \code{E2} and \code{A2}, 
build object instantiations, 
as in section~\ref{ssec:instance}.
The pushout in the middle, with vertex 
\code{Envelope}$\la$\code{Zp}$\ra$,
corresponds to a template parameter passing,
as in section~\ref{ssec:template}.
The horizontal coprojection of the bottom pushout,
from \code{Archetype} to \code{Archetype}~\code{A2(}$\&$\code{E2)},
is composed of three morphisms of different nature:
first an inheritance, then a template parameter passing,
and finally an instantiation.

\begin{figure}[htbp]
\input{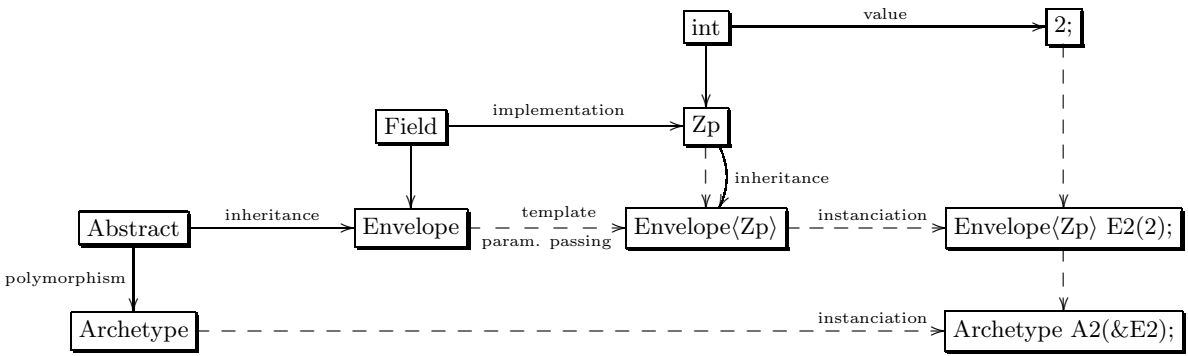}
\caption{A DML diagram for the \linbox architecture for fields,
  with an inheritance envelope}\label{fig:corpsherit}
\end{figure}

In this diagram, the envelope is a copy one:
the construction of the envelope \code{E2}
requires a copy of the field \code{F2}.
This can be compared with the next diagram,
in figure~\ref{fig:corpsherit},
with an inheritance envelope.

It appears clearly now that the difference lies in 
the instantiation of the field. Indeed, the field construction is not
required anymore, it is automatically made during the construction of the
envelope. In figure \ref{fig:corpsherit}, we see that one can still
instantiate \code{Zp(2)} if needed elsewhere, 
but this is actually now automatically done within the
constructor of the envelope.



\section{Conclusion}
We have designed a new diagrammatic modeling language, DML.
The paradigm used is that of the category theory and in particular of
the pushout tool. We have shown that most of the object-oriented
structures can be described with this tool and have many examples in
C++, ranging from virtual inheritance and polymorphism to template
genericity.

With this powerful tool, we propose a quite simple description of the
\linbox library. This library has been designed for efficiency and
genericity and therefore makes heavy usage of complex template and
polymorphic mechanism. Be reverse engineering, we were able to
describe the fundamental structure of archetypes in \linbox.
This structure contains several classes generic or not, polymorphic or
not and our description requires four pushouts. We believe that our
description with pushouts is quite clear and enables better
understanding of the behavior of the archetypes.

Next work will be to have tools to manipulate the diagrams and to
generate object oriented skeletons. Prototypes of such softwares
(``Dessiner les Calculs'' for diagram manipulations and ``SketchUML''
for generating UML for diagrams) are
available on the web page of the InCa
project\footnote{\url{http://www-lmc.imag.fr/MOSAIC/InCa}}.





\begin{thebibliography}{10}

\bibitem{Barr:1990:CTCS}
Michael Barr and Charles Wells.
\newblock {\em {Category Theory for Computer Science}}.
\newblock International Series in Computer Science. Prentice Hall, 1990.

\bibitem{Burstall:1997:PTT}
R.M. Burstall and J.A. Goguen.
\newblock Putting theories together to make specifications.
\newblock In {\em Proc. 5th Internat. Joint Conf. on Artificial Intelligence},
  pages 1045--1058, 1997.

\bibitem{jgd:2002:icms}
Jean-Guillaume Dumas, Thierry Gautier, Mark Giesbrecht, Pascal Giorgi, Bradford
  Hovinen, Erich Kaltofen, B.~David Saunders, Will~J. Turner, and Gilles
  Villard.
\newblock {LinBox}: A generic library for exact linear algebra.
\newblock In Arjeh~M. Cohen, Xiao-Shan Gao, and Nobuki Takayama, editors, {\em
  Proceedings of the 2002 International Congress of Mathematical Software,
  Beijing, China}, pages 40--50. World Scientific Pub, August 2002.

\bibitem{Duval:2002:DS}
D.~Duval and C.~Lair.
\newblock Diagrammatic specifications.
\newblock Rapport de recherche 1043 m, IMAG-LMC, January 2002.

\bibitem{Duval:2003:DS}
Dominique Duval.
\newblock Diagrammatic specifications.
\newblock {\em Mathematical Structures in Computer Science}, 13(6):857--890,
  2003.

\bibitem{Goguen:1973:CFG}
J.A. Goguen.
\newblock Categorical foundations for general systems theory.
\newblock In {\em Advances in Cybernetics and System Research}, pages 121--130.
  Transcripta Books, 1973.

\bibitem{kaltofen:2005:memory}
Erich Kaltofen, Dmitriy Morozov, and George Yuhasz.
\newblock Generic matrix multiplication and memory management in linbox.
\newblock In Manuel Kauers, editor, {\em Proceedings of the 2005 International
  Symposium on Symbolic and Algebraic Computation, Beijing, China}. ACM Press,
  New York, July 2005.

\bibitem{MacLane:1997:CWM}
Saunders {Mac Lane}.
\newblock {\em Categories for the Working Mathematician}, volume~5 of {\em
  Graduate Texts in Mathematics}.
\newblock Springer-Verlag, New York, 2nd edition, 1997.
\newblock (1st ed., 1971).

\bibitem{Muller:2000:UML}
Pierre-Alain Muller and Nathalie Gaertner.
\newblock {\em Mod{\'e}lisation objet avec UML}.
\newblock Eyrolles, 2000.

\bibitem{Oriat:2000:DEM}
Catherine Oriat.
\newblock Detecting equivalence of modular specifications with categorical
  diagrams.
\newblock {\em TCS}, 247(1--2):141--190, 2000.

\bibitem{Srinivas:1995:SPEC}
Y.V. Srinivas and R.~J{\"u}llig.
\newblock Specware language manual, 1995.

\bibitem{Stroustrup:1997:cpp}
Bjarne Stroustrup.
\newblock {\em The {C}++ Programming Language: Third Edition}.
\newblock Addison-Wesley Publishing Co., Reading, Mass., 1997.

\bibitem{Taivalsaari:1996:NI}
Antero Taivalsaari.
\newblock On the notion of inheritance.
\newblock {\em ACM Computing Surveys}, 28(3):438--479, September 1996.

\bibitem{Turner:2002:these}
Will~J. Turner.
\newblock {\em Blackbox linear algebra with the LinBox library}.
\newblock PhD thesis, North Carolina State University, May 2002.

\bibitem{Wise:1996:stl}
G.~Bowden Wise.
\newblock An overview of the standard template library.
\newblock {\em SIGPLAN Not.}, 31(4):4--10, 1996.

\end{thebibliography}
\end{document}